\documentclass[a4paper,12pt]{article}

\usepackage{amsthm,amssymb,amsmath,url,amsfonts,mathrsfs}

\theoremstyle{plain}
\newtheorem{thm}{Theorem}
\newtheorem{lem}{Lemma}
\newtheorem{cor}{Corollary}
\newtheorem{defn}{Definition}

\usepackage{graphicx}
\usepackage{dcolumn}
\usepackage{bm}

\newcommand{\R}{\mathbb{R}}
\newcommand{\X}{\mathbb{X}}

\newcommand{\Z}{\mathbb{Z}}
\newcommand{\G}{\mathbb{G}}

\newcommand{\VVV}{{\bf V}}
\newcommand{\PPP}{{\bf P}}
\newcommand{\QQQ}{{\bf Q}}

\newcommand{\XX}{{\cal X}}
\newcommand{\BB}{{\cal B}}
\newcommand{\HH}{{\cal H}}

\newcommand{\GG}{{\cal G}}
\newcommand{\DD}{{\cal D}}

\newcommand{\TT}{{\cal T}}

\newcommand{\x}{{\bf x}}

\begin{document}
\title{\bf The permanent spatial decomposition of the wave function}
\author{Bruno Galvan \footnote{e-mail: b.galvan@virgilio.it}\\ \small Via di Melta 16, 38121 Trento, Italy.}


\date{\today}
\maketitle
\begin{abstract}
{\it Permanent spatial decomposition} (PSD) is the (hypothesized) property of the wave function of a macroscopic system of decomposing into localized permanently non-overlap\-ping parts when it spreads over a macroscopic region. The typical example of this phenomenon is the measurement process, in which the wave function of the laboratory (quantum system + apparatus + environment) decomposes into $n$ parts, corresponding to the $n$ outcomes of the measurement: the parts are {\it non-overlapping}, because they represent a macroscopic pointer in different positions, and they are {\it permanently} non-overlapping due to the irreversible interaction with the environment. PSD is often mentioned in the literature, but until now no formal definition or systematic study of this phenomenon has been undertaken. The aim of this paper is to partially fill this gap by giving a formal definition of PSD and studying its possible connection with scattering theory. The predictive and explanatory powers of this phenomenon are also discussed and compared with those of Bohmian mechanics.
\end{abstract}


\section{Introduction} \label{introduction}

Here, {\it permanent spatial decomposition} (PSD) of the wave function refers to the (hypothesized) property of the wave function of a macroscopic system of decomposing into localized permanently non-overlapping parts when it spreads over a macroscopic region. The typical example of this phenomenon is the measurement process, in which the wave function of the laboratory (quantum system + apparatus + environment) decomposes into $n$ parts, corresponding to the $n$ outcomes of the measurement: the parts are {\it non-overlapping}, because they represent a macroscopic pointer in different positions, and they are {\it permanently} non-overlapping due to the irreversible interaction with the environment. After decomposition, every part undergoes further spreading and consequent decomposition, and so on, thus determining the emergence of a tree structure for the wave function. The structure of the branches is hypothesized to correspond to the observed quasi-classical macroscopic evolution.

The phenomenon of PSD is an important element in various formulations of quantum mechanics:

(a) In Bohmian mechanics (see \cite{bmbook} and reference therein) it is a commonly accepted fact that the trajectories defined by the guidance equation remain inside the supports of the various non-overlapping parts of the wave function. This fact determines the so-called effective collapse of the wave function: At the end of a measurement the actual trajectory of the laboratory enters into only one of the non-overlapping parts in which the wave function of the laboratory is decomposed; the fact that the ``empty'' parts do not overlap the ``active'' part implies that the evolution of the actual trajectory is not affected by the empty parts in the future, and these can therefore be considered as collapsed. The effective collapse of the wave function guarantees that the predictions of Bohmian mechanics relative to the quantum experiments correspond to those of standard quantum measurement theory, in which the ``empty'' parts are assumed to actually collapse. Most of the studies concerning PSD have been performed in the context of Bohmian mechanics (see \cite{undivuni}, chapters 5 and 6).

(b) According to a formulation of quantum mechanics recently proposed by the author \cite{galvan:4}, the particles follow definite trajectories, even though not necessarily continuous, and a new quantum rule, the {\it Quantum Cournot Principle}, constrains them to remain inside the supports of the non-overlapping parts of the wave function. The phenomenon of PSD is therefore fundamental in this formulation for determining the structure of the trajectories.

(c) According to the Many Worlds Interpretation \cite{everett1, MWI1}, the wave function of the universe splits into branches or {\it worlds}, but the explicit definition of the branches is usually considered a matter of interpretation, and is not included in the mathematical formalism of this formulation. An exception to this approach has been recently proposed in \cite{zanghimw}, where the branches are defined by the criterion of PSD.

In spite of the fact that PSD is often mentioned, no systematic study and not even a formal definition or an explicit name can be found in the literature for this phenomenon. The aim of this paper is to partially fill this gap by proposing the name {\it permanent spatial decomposition} and a formal definition for this phenomenon. A connection between PSD and scattering theory is also argued. More precisely, it is argued that different elements of a PSD belong to different (sets of) scattering channels of the wave function of the universe.

The paper is structured as follows: in section \ref{formal} the basic formalism relative to PSD is developed and a formal definition of PSD is given; in section \ref{localisation} the notion of localized state vector is defined in a general way; in section \ref{examples} an example relative to the previously developed formalism is presented; in section \ref{hypothesis} the hypothesis that the wave function of the universe is subjected to PSD is explicitly and precisely formulated; in section \ref{permanent} a theorem is proved which shows a possible link between PSD and scattering theory; in section \ref{predictive} the predictive and explanatory powers of PSD are discussed and compared with those of Bohmian mechanics; section \ref{summary} is the summary.

\section{Basic formalism and lemmas} \label{formal}

A very rudimentary version of the formalism presented in this section has already been proposed in my previous papers \cite{galvan:1, galvan:4}. Even though the basic idea is the same, the formalism developed here is new, simpler and more rigorous compared with those presented in the papers cited above. Most of the lemmas are straightforward and their proof is omitted.

The basic mathematical elements are a Hilbert space $\cal H$, a self-adjoint Hamiltonian $H$, which defines the group of unitary time evolution operators $U(t)=\exp[-i H t/\hbar]$, and a generic spectral measure $G$ defined on a metric space $\G$. The evolution will always be considered along the positive time axis $[0, +\infty)$. For a formal definition and various properties of spectral measures, see the appendix. The symbol $E$ instead of $G$ will be used to denote the spatial spectral measure on configuration space, and configuration space will be denoted by $\X$.

\begin{defn} 
A decomposition of a vector $\Psi \in {\cal H}$ is a finite unordered set of linearly independent vectors $\{\Psi_1, \ldots, \Psi_n\}$ such that $\sum_i \Psi_i=\Psi$; if the vectors are pairwise orthogonal, the decomposition is said to be orthogonal.
\end{defn}

Let us introduce the following notations: $\DD$ denotes the decomposition $\{\Psi_1, \ldots, \Psi_n\}$, $\#\DD$ is the cardinality of $\DD$, $\sum \DD$ is the vector $\sum_i \Psi_i$, and $U(t)\DD$ is the decomposition $\{U(t)\Psi_1, \ldots, U(t)\Psi_n\}$. An orthogonal decomposition is naturally endowed with a probability measure $P$ defined as follows: $P(\Psi_i):=||\Psi_i||^2/||\sum \DD||^2$. This definition corresponds to the Born rule.

\begin{defn}
A decomposition $\DD$ is said to be finer than a decomposition $\DD'$ (or equivalently, $\DD'$ is said to be coarser than $\DD$) if there exists a map $h:\DD \to \DD'$ such that $\Psi'_i=\sum h^{-1}[\Psi'_i]$ for any $\Psi'_i \in \DD'$.
\end{defn}
If $\DD$ if finer than $\DD'$, we write $\DD \preceq \DD'$. In words, $\DD \preceq \DD'$ if every element of $\DD'$ is the sum of a different subset of elements of $\DD$. 
\begin{lem}
(a) if the map $h$ exists, it is univocally defined and surjective; (b) $\preceq $ is a partial order (c) $\DD \preceq  \DD'$ implies that $\sum \DD= \sum \DD'$ and $U(t)\DD \preceq  U(t)\DD'$.
\end{lem}
If $\DD \preceq \DD'$, $\Psi_i \in \DD$, $\Psi'_j \in \DD'$ and $\Psi_i \in h^{-1}[\Psi'_j]$, we write $\Psi_i \subseteq \Psi'_j$.

The following notion of {\it tree} formalizes the situation in which every element of decomposition undergoes subsequent decompositions.
\begin{defn}
A (forward) tree $\TT$ for the vector $\Psi_0$ is a finite sequence $\{(t_1, \DD_1), \ldots, (t_r, \DD_r)\}$, where:
\\(a) $\{t_1, \ldots, t_r\}$ is a sequence of times with $0 \leq t_1 < t_2 < \ldots < t_r$;
\\(b) $\{\DD_1, \ldots, \DD_r\}$ is a sequence of decompositions such that $\sum \DD_1=U(t_1)\Psi_0$ and
\begin{equation}
\DD_{i+1} \preceq U(t_{i+1}-t_i)\DD_i \; \hbox{ for } \; i=1, \ldots,r-1.
\end{equation}
\end{defn}
Note that from these properties it follows that $\sum \DD_i=U(t_i)\Psi_0$ for $1 \leq i \leq r$.

\begin{lem}
If the last decomposition of a tree is orthogonal then all the decompositions of the tree are orthogonal.
\end{lem}
A tree naturally defines a time-dependent decomposition:
\begin{defn}
Let $\TT=\{(t_1, \DD_1), \ldots, (t_r, \DD_r)\}$ be a tree for the vector $\Psi_0$. The map $\hat \TT$ from the time interval $[0, \infty)$ with values in the set of the decompositions is defined as follows: \\
\begin{equation}
\hat \TT(t):= \left \{
\begin{array}{ll}
U(t)\{\Psi_0\} & \hbox{ for } t \in [0, t_1);\\
U(t-t_i)\DD_i & \hbox{ for } t \in [t_i, t_{i+1}), \; i=1, \ldots,r-1; \\
U(t-t_r)\DD_r & \hbox{ for }t \geq t_r.
\end{array}\right.
\end{equation}
\end{defn}
\begin{lem}
(a) $\hat \TT(t_2) \preceq  U(t_2-t_1) \hat \TT(t_1)$ for $t_2 \geq t_1$; (b) $\lim_{\epsilon \to 0^+} \#\hat\TT(t+\epsilon)=\#\hat\TT(t)$.
\end{lem}
\begin{defn}
A branch of a tree $\TT$ is a map $\Phi:[0, \infty) \to \HH$ satisfying the following properties: (a) $\Phi(t) \in \hat \TT(t)$ for any $t \geq 0$; (b) $\Phi(t_2) \subseteq U(t_2-t_1)\Phi(t_1)$ for $t_2 \geq t_1$.
\end{defn}
\begin{lem}
(a) If $\Phi_1(t)=\Phi_2(t)$ for a given $t$ and two branches $\Phi_1$ and $\Phi_2$ of a tree, then $\Phi_1(s)=\Phi_2(s)$ for $s \leq t$; (b) for every $\Psi_i$ belonging to the last decomposition $\DD_r$ of a tree, there is exactly one branch $\Phi_i$ such that $\Phi_i(t_r)=\Psi_i$.
\end{lem}
From point (b) of the previous lemma, it follows that there is a one-to-one correspondence between the branches of the tree and the elements of the last decomposition.

\vspace{3mm}
The above formalism defines in a general way the notions of decomposition, tree and branches of a wave function, and arguably it is valid whatsoever criterion is utilized for decomposing the wave function. Let us now develop a formalism that will be useful for defining spatial decompositions.

Let $\GG:=\{\Delta_1, \ldots , \Delta_n\}$ be a finite partition of $\G$, and let $G(\GG)\Psi$ denote the decomposition $\{G(\Delta_1)\Psi, \ldots, G(\Delta_n)\Psi\}$.
\begin{defn}
A decomposition $\DD$ is said to be an exact $G$-decomposition if there exists a partition $\GG$ of $\G$ such that $\DD=G(\GG) \Psi$, where $\Psi=\sum \DD$.
\end{defn}
Of course an exact $G$-decomposition is an orthogonal decomposition. Hereafter, when $G$ is the spatial spectral measure $E$, the prefix $E$- will be replaced by the word ``spatial''; thus an exact $E$-decomposition will be referred to as an exact spatial decomposition.

We now want to find a function which measures the ``distance'' of a generic decomposition from an exact $G$-decomposition. Let us consider the following tentative definition:
\begin{equation} \label{e1}
w_G(\DD)\stackrel{\hbox{\tiny tentative}}{:=}\inf_{\GG} \max_{1 \leq i \leq n} \left \{\frac{||\Psi_i - G(\Delta_i)\Psi||}{||\Psi_i||} \right \},
\end{equation}
where $n = \#\DD$, $\Psi=\sum \DD$, and $\GG$ ranges over the partitions of $\G$ with $n$ elements. We could therefore say that a decomposition $\DD$ is approximately a $G$-decomposition if $w_G(\DD) \approx 0$. A natural property to require for the function $w_G$ is that if $w_G(\DD) \approx 0$ and $\DD'$ is coarser than $\DD$, then $w_G(\DD') \approx 0$ also holds true. This requirement suggests that we modify the above definition as follows: if $I$ is a subset of $\{1, \ldots, n\}$, define $\Psi_I:=\sum_{i \in I}\Psi_i$ and $\Delta_I:=\cup_{i \in I} \Delta_i$. The definition (\ref{e1}) is modified as follows:
\begin{defn}
The function $w_G$ is defined as:
\begin{equation}
w_G(\DD):=\inf_{\GG} \max_I \left \{\frac{||\Psi_I - G(\Delta_I)\Psi||}{||\Psi_I||} \right \}.
\end{equation}
If $w_G(\DD) \approx 0$ the decomposition $\DD$ is said to be a (approximate) $G$-decomposition.
\end{defn}
Hereafter, the attribute {\it approximate} will be omitted. The following lemma guarantees that the function $w_G$ has the required property:
\begin{lem}
$\DD \preceq  \DD'$ implies that $w_G(\DD') \leq w_G(\DD)$.
\end{lem}

{\it Example.} Let us study the function $w_G$ for a two-state decomposition $\{\Psi_1, \Psi_2\}$. Since
\begin{eqnarray*}
& & ||\Psi_1 - G(\Delta)\Psi||=||\Psi_2 - G(\Delta^c)\Psi|| \\
& & =\left (||G(\Delta^c) \Psi_1||^2 + ||G(\Delta) \Psi_2||^2 \right )^{1/2},
\end{eqnarray*}
we obtain
\begin{equation} \label{e4}
w_G(\{\Psi_1, \Psi_2\}) = \inf_{\Delta}\frac{\left(||G(\Delta^c) \Psi_1||^2 + ||G(\Delta) \Psi_2||^2\right )^{1/2}}{\min\{||\Psi_1||, ||\Psi_2||\}}.
\end{equation}
Consider the measures $\mu_1:=||G(\cdot) \Psi_1||^2$, $\mu_2:=||G(\cdot) \Psi_2||^2$, and the signed measure $\mu:=\mu_1 - \mu_2$. According to the Hahn decomposition theorem, there exists a measurable set $\tilde \Delta$  such that $\mu_1(\Delta) \geq \mu_2(\Delta)$ for $\Delta \subseteq \tilde \Delta$ and $\mu_1(\Delta) \leq \mu_2(\Delta)$ for $\Delta \subseteq \tilde \Delta^c$. As a consequence, equation (\ref{e4}) becomes:
\begin{equation}
w_G(\{\Psi_1, \Psi_2\}) = \frac{\left(||G(\tilde \Delta^c) \Psi_1||^2 + ||G(\tilde \Delta) \Psi_2||^2\right )^{1/2}}{\min\{||\Psi_1||, ||\Psi_2||\}}.
\end{equation}
From this equation one easily obtains the results: $0 \leq w_G(\{\Psi_1, \Psi_2\}) \leq 1$; $w_G(\{\Psi_1, \Psi_2\}) = 0 $ iff $\mu_1$ and $\mu_2$ have disjoint supports, and $w_G(\{\Psi_1, \Psi_2\}) = 1 $ iff $\mu_1(\Delta) \leq \mu_2(\Delta)$ or $\mu_1(\Delta) \geq \mu_2(\Delta)$ for any measurable $\Delta$. For $G=E$ we can write
\begin{equation} \label{17}
w_E(\{\Psi_1, \Psi_2\})=\frac{\left(\int \min\{|\Psi_1(x)|^2, |\Psi_2(x)|^2\} dx\right)^{1/2}} {\min\{||\Psi_1||, ||\Psi_2||\}},
\end{equation}
where $\Psi_i(x)$, for $i=1, 2$, is the state vector in the coordinate representation. An analogous expression can be obtained for the momentum spectral measure. The above expressions was already presented in \cite{galvan:1}.

The following lemma ensures that, to a very good approximation, a $G$-decomposition is an orthogonal decomposition.
\begin{lem} \label{scal}
For $I, J \subseteq \{1, \ldots, n\}$, with $J \cap K=\emptyset$, we have
\begin{equation}
\frac{|\langle \Psi_I | \Psi_J \rangle|}{||\Psi_I|| ||\Psi_J||} \leq 2 w_G(\DD) + w^2_G(\DD).
\end{equation}
\end{lem}
\begin{proof}
For any $\epsilon >0$ there exists a partition $\{\Delta_1, \ldots, \Delta_n\}$ of $\G$ such that, for any $I \subseteq \{1, \ldots, n\}$ we have
$$
\frac{||\Psi_I - G(\Delta_I)\Psi||}{||\Psi_I||} \leq w_G(\DD) + \epsilon,
$$
where. as usual, $\Psi=\sum \DD$. Then
\begin{eqnarray*}
& & \frac{|\langle \Psi_I | \Psi_J \rangle|}{||\Psi_I|| ||\Psi_J||}=\frac{|\langle \Psi_I -G(\Delta_I)\Psi + G(\Delta_I)\Psi| \Psi_J \rangle|}{||\Psi_I|| ||\Psi_J||} \\
& & \leq \frac{||\Psi_I -G(\Delta_I)\Psi|| ||\Psi_J|| + |\langle G(\Delta_I)\Psi|\Psi_J - G(\Delta_J)\Psi + G(\Delta_J)\Psi \rangle|}{||\Psi_I|| ||\Psi_J||} \\
& & \leq w_G(\DD) + \epsilon + \frac{||G(\Delta_I)\Psi||||\Psi_J- G(\Delta_J)\Psi||}{||\Psi_I|| ||\Psi_J||} \\
& & \leq w_G(\DD) + \epsilon + [w_G(\DD) + \epsilon] \frac{||\Psi_I|| + ||\Psi_I- G(\Delta_I)\Psi||}{||\Psi_I||} \\
& & \leq w_G(\DD) + \epsilon + [w_G(\DD) + \epsilon][1+ w_G(\DD) + \epsilon].
\end{eqnarray*}
\end{proof}
\begin{thm}
$w_G(\DD)=0$ iff $\DD$ is an exact $G$-decomposition.
\end{thm}
\begin{proof} The implication ``$\DD$ is an exact $G$-decomposition'' $\Rightarrow w_G(\DD)=0$ is obvious. If $w_G(\DD)=0$ then $\inf_\Delta ||\Psi_i - G(\Delta)\Psi||=0$ for $i=1, \ldots, n$. From the previous example we deduce that for every $i$ there exists a set $\Delta_i$ such that $\Psi_i=G(\Delta_i)\Psi$. We define $\Sigma_i:=\Delta_i \cap_{j \ne i} \Delta_j^c$ for $i=1, \ldots, n$, and $\Sigma_0:=\X \setminus (\cup_i  \Sigma_i)$. Then $\GG:=\{\Sigma_0 \cup \Sigma_1, \Sigma_2, \ldots, \Sigma_n \}$ is a partition of $\G$. From lemma \ref{scal} we have $0=\langle \Psi_i|\Psi_j \rangle = \langle \Psi_i |G(\Delta_j)|\Psi_i \rangle$ for $i \neq j$, from which $G(\Delta^c_j)\Psi_i=\Psi_i$. Thus $G(\Sigma_i)\Psi=\Psi_i$ for $i=1, \ldots, n$, $G(\Sigma_0)\Psi=\Psi - \sum_i \Psi_i=0$, and therefore $\DD=G(\GG)\Psi$.
\end{proof}
A $G$-decomposition can be naturally endowed with a probability measure which very precisely approximates the Born rule. In fact, if $w_G(\DD) \approx 0$, there exists a partition $\{\Delta_1, \ldots, \Delta_n\}$ of $\G$ such that
$$
\frac{||\Psi_i - G(\Delta_i)\Psi||}{||\Psi_i||} \approx 0 \hbox{ for } i=1, \ldots, n.
$$
We can therefore define $P(\Psi_i)=||G(\Delta_i)\Psi||^2/||\Psi||^2 \approx ||\Psi_i||^2/||\Psi||^2$.

In order to express the fact that a decomposition is a $G$-decomposition in a permanent way let us introduce the function $w_G^+$:
\begin{defn}
The function $ w_G^+$ is defined as follows:
\begin{equation}
w_G^+(\DD):=\sup_{t \geq 0} w_G[U(t)\DD].
\end{equation}
If $w_G^+(\DD) \approx 0$, the decomposition $\DD$ is said to be a permanent $G$-decomposition.
\end{defn}
According to the previously introduced convention, a permanent $E$-decomposition will be referred to as a permanent spatial decomposition (PSD). The notion of PSD formally defines of the phenomenon of decomposition of the wave function into permanent non-overlapping parts which has been described in the introduction. 

{\it Example}. Let us explicitly write down the condition for which the decomposition $U(t)E(\XX)\Psi$, where $\XX$ is a partition of the configuration space, is still a spatial decomposition:
\begin{equation} \label{09}
w_E[U(t)E(\XX)\Psi]=\inf_{\XX'} \max_I \frac{||E(\Delta'_I)U(t)\Psi - U(t)E(\Delta_I) \Psi||}{||E(\Delta_I)\Psi||} \approx 0.
\end{equation}
The above expression relates the formalism developed here to that presented in \cite{galvan:4}. In fact, the condition (\ref{09}) is a generalization of the condition defining the branches of the tree structure in \cite{galvan:4}; see equations (8), (9), (10), (17), and (38) in that paper.

\begin{lem} 
If $G$ commutes with the Hamiltonian, then $w_G(\DD)=w_G[U(t)\DD]=w_G^+(\DD)$.
\end{lem}

It is natural to extend the function $w_G^+$ to trees:
\begin{defn}
The function $w_G^+$ for a tree $\TT$ is defined as follows:
\begin{equation}
w_G^+(\TT):= \sup_{t \geq 0} w_G[\hat \TT(t)].
\end{equation}
If $w_G^+(\TT) \approx 0$, the tree $\TT$ is said to be a (approximate) $G$-tree.
\end{defn}
\begin{lem}
Let $\TT$ be the tree $\{(t_1, \DD_1), \ldots, (t_r, \DD_r)\}$; then: (a) $w_G^+(\TT)=\max\{w_G^+(\DD_1), \ldots, w_G^+(\DD_r)\}$; (b) if $G$ commutes with the Hamiltonian then $w_G^+(\TT)=w_G(\DD_r)$.
\end{lem}
The last decomposition of a $G$-tree is a $G$-decomposition and can therefore be naturally endowed with a probability measure. Since there is a one-to-one correspondence between the last decomposition and the set of the branches of a tree, such a set can also be naturally endowed with a probability measure.

\section{Localization of state vectors} \label{localisation}

The decomposition of a wave function into permanently non-overlapping branches is meaningful from the physical point of view if the branches are localized on a macroscopic scale. In this section, the notion of localized wave function will be defined in a general way.

Let us consider first the case of a single particle. In this case, the configuration space is $\R^3$. We define the {\it centroid} \, ${\bf x}_\Psi \in \R^3$ of a vector $\Psi$ as follows:
\begin{equation} \label{e2}
{\bf x}_\Psi:=\frac{\int {\bf y} \langle \Psi | dE({\bf y})|\Psi \rangle }{\langle \Psi |\Psi \rangle}.
\end{equation}
The spread $\sigma_\Psi \in \R$ of $\Psi$ is defined in the usual way:

\begin{equation} \label{e3}
\sigma^2_\Psi:= \frac{\int ({\bf y} - {\bf x}_\Psi)^2 \langle \Psi | dE({\bf y})|\Psi \rangle }{\langle \Psi |\Psi \rangle}.
\end{equation}
We therefore say that $\Psi$ is {\it localized} if $\sigma_\Psi$ is small on a macroscopic scale.

The definitions of centroid and spread can be generalized to a generic metric configuration space $\X$.

For any $x \in \X$ define:
\begin{equation}
\tau^2_\Psi(x):= \frac{\int d^2(x, y) \langle \Psi | dE(y)|\Psi \rangle }{\langle \Psi |\Psi \rangle},
\end{equation}
where $d$ is the distance on $\X$. Then we define:
\begin{equation}
\sigma_\Psi:=\inf_{x \in \X} \tau_\Psi(x).
\end{equation}
We will assume that in all cases of physical interest the lower bound is a minimum, and that a single configuration $x_\Psi$ exists such that $\tau_\Psi(x_\Psi)=\sigma_\Psi$. In this case $x_\Psi$ is the centroid. It is easy to see that we obtain the centroid (\ref{e2}) and the spread (\ref{e3}) if the distance on $\R^3$ is defined as $||{\bf x} - {\bf y}||$.

For a system of $N$ particles of mass $m_1, \ldots, m_N$ the configuration space $\R^{3N}$ is naturally endowed with the scalar product 
\begin{equation} \label{scap}
\langle x| y \rangle :=\frac{\sum_i m_i {\bf x_i} \cdot {\bf y_i}}{M},
\end{equation}
where $M=\sum m_i$. The scalar product (\ref{scap}) defines the distance
\begin{equation}
d(x, y)=\sqrt{ \frac{\sum_i m_i ||{\bf x}_i - {\bf y}_i||^2}{M}}.
\end{equation}
This definition satisfies the natural requirement that two particles of mass $m_i$ and $m_j$ in the same position $\bf x$ can be considered as a single particle of mass $m_i + m_j$. With this definition we obtain
\begin{equation}
x_\Psi:=\frac{\int y \langle \Psi | dE(y)|\Psi \rangle }{\langle \Psi |\Psi \rangle}
\end{equation}
and
\begin{equation}
\sigma^2_\Psi:= \frac{\int \sum_i m_i ({\bf y}_i - {\bf x}_{\Psi i})^2 \langle \Psi | dE(y)|\Psi \rangle }{M \langle \Psi |\Psi \rangle}.
\end{equation}

The definition of centroid allows us to associate a trajectory $\gamma_\Phi:[0, \infty) \to \X$ with every branch $\Phi$ of a tree:
\begin{equation}
\gamma_\Phi(t):=x_{\Phi(t)}.
\end{equation}
Since the set of branches of a spatial tree is naturally endowed with a probability measure, a set of trajectories endowed with a probability measure can be associated with a spatial tree. This fact will be utilized in section \ref{predictive}, where the predictive power of PSD will be discussed.

\section{Example} \label{examples}

Let us consider a one-dimensional free particle of mass $m$, whose wave function is the sum of two Gaussian wave packets traveling in opposite directions. Initially, the two wave packets are located at the origin; after a suitable time they cease to overlap and the wave function can be spatially decomposed in a permanent way.

Let $\Psi_0=\Psi_- + \Psi_+$, where, in the momentum representation, 
\begin{equation}
\Psi_\pm(p) =\left (\frac{1}{\sqrt{\pi} \sigma_p} \right)^{1/2}\exp \left (- \frac{(p \mp p_0)^2}{2 \sigma_p^2}\right),
\end{equation}
with $p_0 >0$. Let $F$ denote the momentum spectral measure. We have 
\begin{equation} 
w_F(\DD) =\left(\int \min\{|\Psi_-(p)|^2, |\Psi_+(p)|^2\} dp\right)^{1/2},
\end{equation}
where $\DD:=\{\Psi_-, \Psi_+\}$. Since $|\Psi_-(p)|^2 \leq |\Psi_+(p)|^2$ for $p \in \R^+$, and from symmetry considerations, we obtain:
\begin{eqnarray}
& & w_F(\DD)= \frac{2}{\sqrt{\pi} \sigma_p} \int_0^\infty \exp \left (- \frac{(p + p_0)^2}{\sigma_p^2}\right)dp \\
& & = \frac{2}{\sqrt{\pi}} \int_{p_0/\sigma_p}^\infty e^{-y^2} dy
= \hbox{erfc}(p_0/\sigma_p), \nonumber
\end{eqnarray}
where $\hbox{erfc}()$ is the complementary error function. We assume that $ w_F(\DD) \ll1$. For example, for $p_0/\sigma_p=10$ we have $ w_F(\DD)=4.6 \times 10^{-23}$. The elements of $\DD$ are orthogonal with very high precision, and the decomposition can be endowed with the probability measure $P(\Psi_\pm)=1/2$.

Let us now calculate $w_E[U(t)\DD]$. By applying the Schr\"{o}dinger equation, we obtain the well known formula for the evolution of the square modulus of Gaussian wave packets:
\begin{equation} \label{20}
|U(t)\Psi_\pm|^2(x)=\frac{1}{\sqrt{\pi} \sigma(t)}\exp \left (- \frac{[x \mp x(t)]^2} {\sigma^2(t)}\right ),
\end{equation}
where 
\begin{equation} \label{21}
x(t)= tp_0/m, \; \; \sigma(t)=\sqrt{\sigma_x^2 + \frac{\hbar^2 t^2}{ m^2 \sigma_x^2 }}, \hbox{ and } \sigma_x=\frac{\hbar}{\sigma_p}.
\end{equation}
The calculation for obtaining $w_E[U(t) \DD]$ is analogous to that for obtaining $w_F[\DD]$. Thus, we have:
\begin{equation} \label{22}
w_E[U(t) \DD]= \hbox{erfc}[x(t)/\sigma(t)].
\end{equation}
Let us study the function
\begin{equation} \label{23}
f(t):=\frac{x(t)}{\sigma(t)}=\frac{p_0 }{\sigma_p}\frac{t} {\sqrt{ m^2 \hbar^2/\sigma_p^4 + t^2}}
\end{equation}
for $t \geq 0$. We then have: $f(0)=0$; the derivative is never zero and, since $f'(0) > 0$, we have $f'(t) > 0$ for $t > 0$; moreover $\lim_{t \to \infty}f(t)=p_0/\sigma_p$. Thus, $f(t)$ increases from the value $0$ at the time $t=0$ to the asymptotic limit $p_0/\sigma_p$. As a consequence, $w_E[U(t) \DD]$ decreases from the value $1$ at the time $t=0$, when the two wave packets are totally overlapping, to the asymptotic value $\hbox{erfc}(p_0/\sigma_p)=w_F(\DD)$, when the spatial overlapping of the wave packets is equal to their overlapping in momentum space. As a consequence
\begin{equation}
w_E^+[U(t)\DD]=w_E[U(t)\DD].
\end{equation}
In order to define a spatial tree for $\Psi_0$ it is sufficient to choose a time $t_1$ such that $w_E[U(t_1) \DD] \approx 0$. The tree will be 
\begin{equation}
\{(t_1, U(t_1)\DD)\}.
\end{equation}
Of course, the time $t_1$ is only vaguely defined. The branches $\Phi_\pm(t)$ are:
\begin{equation}
\Phi_\pm(t)= \left \{
\begin{array}{ll}
U(t)\Psi_0 & \hbox{ for } 0 \leq t <t_1;\\
U(t)\Psi_\pm & \hbox{ for } t \geq t_1,
\end{array}\right.
\end{equation}
and the corresponding trajectories $\gamma_\pm(t)$ are: 
\begin{equation}
\gamma_\pm(t)= \left \{
\begin{array}{ll}
0 & \hbox{ for } 0 \leq t <t_1;\\
\pm t p_0/m & \hbox{ for } t \geq t_1.
\end{array}\right.
\end{equation}
The trajectories have a discontinuity at $t_1$. This is not a problem because, as we will see, the trajectories associated with a spatial tree are better interpreted as tools for expressing the predictive power of the wave function rather than the ontological trajectories followed by the particle.

\section{The hypothesis of permanent spatial decomposition} \label{hypothesis}

The formalism developed in the previous sections allows us to formulate in a more precise way the hypothesis assumed in a more or less explicit way in the formulations of quantum mechanics presented in the introduction.

\vspace{3mm}
{\it The PSD hypothesis}: the wave function of the universe admits a spatial tree whose typical branches (i.e., the overwhelming majority of the branches, weighted with their probabilities) are localized in a suitable time interval, which includes the present age of the universe and possibly excludes the initial age close to the big-bang and the final asymptotic age. The tree is only vaguely defined, but vagueness disappears on a macroscopic scale. The set of trajectories defined by the tree correctly predicts the results of the statistical experiments and the structure of macroscopic evolution, according to a criterion that will be better explained in section \ref{predictive}.

\vspace{3mm}
Two different strategies can arguably be adopted in order to prove the PSD hypothesis: (i) to study specific physical processes in which PSD occurs and (ii) to provide structural reasons for which the wave function of the universe would have to be subjected to PSD. Strategy (i) has been adopted for example by Bohm and Hiley \cite{undivuni}. A reasoning of type (ii) will be proposed in the next section, where a connection between PSD and scattering theory is argued.

\section{Permanent spatial decomposition and scattering theory} \label{permanent}

In the literature, the elements of a PSD are usually assumed to be non-overlapping for a very long time, but not permanently. This is due to Poincar\'e recurrence, i.e., to the fact that, for any $\epsilon >0$, there exists a time $T$ such that $||U(T)\Psi_0 - \Psi_0|| \leq \epsilon$ \cite{bmbook}. However, Poincar\'e recurrence only takes place if $\Psi_0$ is a bound state, and modern cosmology suggests on the contrary that the universe is indefinitely expanding. This fact prevents the universe to be in a bound state and allows us to take the attribute {\it permanent} in the definition of PSD seriously. This leads naturally to a correlation of PSD with scattering theory. More specifically, a reasonable working hypothesis is that the non-overlapping parts in which the wave function decomposes correspond to different (sets of) scattering channels of wave function of the universe. In this section a theorem in support of this line of reasoning will be presented.

\vspace{3mm}
Let us consider an $N$-particle system with position operators $Q=(\QQQ_1, \ldots, \QQQ_N)$ and momentum operators $P=(\PPP_1, \ldots, \PPP_N)$. The configuration space $\X$ is $\R^{3N}$. The Hamiltonian of the system is of the type 
\begin{equation} \label{28}
H=\sum_{i=1}^N\frac{\PPP^2_i}{2m_i} + \sum _{i < j} V_{i, j}(\QQQ_i - \QQQ_j),
\end{equation}

Assuming that the potentials $V_{i, j}$ satisfy suitable conditions at infinity (which include Coulomb potentials)\footnote{See \cite{derez}, theorem 6.6.1 for the exact condition; see also the remark at the beginning of section 6.6.}, one can prove the following theorem:
\begin{thm}
The limit 
\begin{equation} \label{limit}
s-C_\infty-\lim_{t \to \infty} \frac{U(-t)QU(t)}{t}=:V_+
\end{equation}
exists, and $V_+$ is a vector of self-adjoint commuting operators which also commute with the Hamiltonian.
\end{thm}
\begin{proof}
See \cite{derez}, theorem 6.6.1.
\end{proof}
Note that the limit (\ref{limit}) is not the usual strong limit of operators, but a kind of limit which is more appropriate for vector operators, and it is equivalent to the weak limit of the spectral measures defined by the vector operators. See the appendix.

The vector operator $V_+$ is referred to as the {\it asymptotic velocity}. For example, for a free particle of mass $m$ we have $\VVV_+=\PPP/m$, and for a particle subjected to a potential admitting the wave operator $\Omega_+$ we have $\VVV_+=\Omega_+ \PPP \Omega_+^\dagger$ (see \cite{derez}, theorem 4.6.1). However the asymptotic velocity exists also for Coulomb potentials, for which wave operator does not exists.

Let $F_+$ denote the spectral measure defined by $V_+$, which of course commutes with the Hamiltonian. Such a measure defines the scattering subspaces of the various channels of the wave function, as I will now explain. A {\it channel} is a partition 
\begin{equation}
a=\{C_1, \ldots, C_r\}
\end{equation}
of the set of particles. The subsets of the channel are referred to as {\it clusters}. The scattering subspace of the Hilbert space corresponding to a channel is composed of the states of the system in which the particles of every cluster are bound together asymptotically and the various clusters move away freely. For every channel $a$ let us define the following subset of configuration space:
\begin{equation}
\Z_a:=\{x \in \X: \x_i=\x_j \hbox{ iff } i, j \in C_k \hbox{ for some } C_k \in a\}
\end{equation}
The sets $\Z_a$ form a partition of $\X$. The subspace $F_+(\Z_a)\HH$ is the subspace of the channel $a$. See \cite{derez} once again for a more detailed explanation.

The following theorem is the new result presented in this section:
\begin{thm} \label{tbase}
$\lim_{t\to \infty} w_E[U(t)F_+(\XX)\Psi] = 0$ for every vector $\Psi \in \HH$ and any partition $\XX$ of $\X$.
\end{thm}
\begin{proof}
Let $\XX=\{\Delta_1, \ldots, \Delta_n\}$: According to corollary \ref{cor1} in the appendix, for any $\epsilon > 0$ there exists a partition $\{\Sigma_1, \ldots, \Sigma_n\}$ such that $F_+(\partial \Sigma_i) = 0$ for $i=1, \ldots, n$ and $||F_+(\Delta_I)\Psi - F_+(\Sigma_I)\Psi|| \leq \epsilon ||F_+(\Delta_I)\Psi||$ for any $I \subseteq \{1, \ldots, n\}$. If $F_t$ denotes the spectral measure defined by $U(-t)QU(t)/t$, we have $F_t(\Delta)=U(-t)E(t \Delta) U(t)$, where $t \Delta:=\{t x \in \X: x \in \Delta\}$. According to lemma \ref{lem1} in the appendix, $F_t(\Sigma_I)\Psi \to F_+(\Sigma_I)\Psi$ for $t \to \infty$. Let $\XX'=\{\Delta'_1, \ldots, \Delta'_n\}$ denote a generic partition of $\X$. We have:
\begin{eqnarray*}
& & w_E[U(t)F_+(\XX)\Psi] = \inf_{\XX'} \max_I \{||F_+(\Delta_I)\Psi - U(-t)E(\Delta'_I)U(t)\Psi||/||F_+(\Delta_I)\Psi||\} \\
& & \leq \inf_{\XX'} \max_I \{||F_+(\Sigma_I)\Psi - U(-t)E(\Delta'_I)U(t)\Psi||/||F_+(\Delta_I)\Psi||\}  \\
& & + \max_I\{||F_+(\Sigma_I)\Psi - F_+(\Delta_I)\Psi||/||F_+(\Delta_I)\Psi||\} \\
& & \leq \max_I \{||F_+(\Sigma_I)\Psi - U(-t)E(t \Sigma_I)U(t)\Psi||/||F_+(\Delta_I)\Psi||\} + \epsilon \to \epsilon \hbox{ for } t \to \infty.
\end{eqnarray*}
Since $\epsilon$ is arbitrary, this proves the theorem.
\end{proof}
This theorem shows that, at least in the asymptotic regime, the wave function has a structural tendency to decompose into permanently non-overlapping parts. More specifically, the theorem guarantees that states corresponding to different (sets of) channel subspaces eventually become non-overlapping (note that states belonging to the same channel subspace but to disjoint subsets of asymptotic velocities also eventually become non-overlapping). A possible strategy to prove the PSD hypothesis is to show that the spatial separation of channels begins long before the wave function enters the asymptotic regime.

According to this reasoning, the PSD hypothesis presented in the previous section can be strengthened as follows:

{\it The asymptotic PSD hypothesis}: the wave function of the universe admits a spatial tree with the properties described by the PSD hypothesis of section \ref{hypothesis}, and moreover its decompositions are of the form $U(t_i)F_+(\XX_i)\Psi_0$.

Note that in this case, due to the defining condition of trees, the partitions $\XX_i$ must satisfy $\XX_i \preceq \XX_j$ for $i \geq j$.

\section{Predictive and explanatory power of permanent spatial decomposition} \label{predictive}

Here, the {\it predictive power} of a physical theory refers to the capacity of the theory to make empirically verifiable predictions\footnote{I utilize here the term {\it prediction} even though the predicted phenomena are already known.}, while {\it explanatory power} refers to the capacity of the theory to explain in a more general sense the nature of physical reality and the origin of observed phenomena.

In the formulations of quantum mechanics mentioned in the introduction, PSD is associated with other laws, for example the guidance equation of Bohmian mechanics or the Quantum Cournot Principle, which play the principal role in those formulations. In this section I will argue that PSD has an autonomous {\it predictive} power, which does not require the presence of other laws. Nevertheless, these laws confer an {\it explanatory} power on the corresponding formulations, explanatory power that PSD alone cannot provide. The reasoning will be based on the assumption that the PSD hypothesis is correct. In what follows, the PSD hypothesis and the related formalism will be provisionally referred to as PSD theory.

The predictive power of PSD theory, analogously to that of Bohmian mechanics, is based on the predictive paradigm of what I have called {\it path spaces}, i.e., sets of trajectories endowed with a probability measure \cite{galvan:1}. The typical example of path space is a canonical stochastic process $(\X^T, {\cal F}, P)$, where $\X^T$ is the set of all the trajectories from a suitable time interval $T$ to a configuration space $\X$, $\cal F$ is the $\sigma$-algebra generated by the cylinder sets and $P$ is the probability measure. Bohmian mechanics also defines a path space: the paths are those satisfying the guidance equation, and the measure of a subset $\Sigma$ of paths is $||E(\Sigma_t)\Psi(t)||^2$, where $\Sigma_t:=\{\lambda(t) \in \X: \lambda \in \Sigma\}$. The equivariance property of the guidance equation guarantees that this definition does not depend on the time. The path space derived from a wave function $\Psi$ according to Bohmian mechanics will be denoted by $\mathscr{B}_\Psi$. We have seen in the previous sections that PSD also allows us to derive a path space from a wave function, though in a vague way. Such a path space will be denoted by $\mathscr{D}_\Psi$.

The predictive paradigm of path spaces consists of assuming that the evolution of the universe is represented by a trajectory chosen at random from a suitable path space and stating that a law of evolution is predicted by the path space if the probability of the set of trajectories satisfying that law is very close to 1. This statement is based on Cournot's principle, which, in the context of probability theory, establishes that an event singled out in advance with probability very close to 1 will happen with empirical certainty in a single trial of a statistical experiment. Cournot's principle is at the base of the predictive power of probability theory; see \cite{galvan:4} and references therein for further information about Cournot's principle. The predictive paradigm of path spaces was first introduced and studied in connection with Bohmian mechanics \cite{qequilibrium}, even though the original idea probably dates back to Boltzmann \cite{seldonboltz}.

Path spaces predict the macroscopic evolution and the results of the statistical experiments in a unified way. Roughly speaking, the former is predicted by the structure of the trajectories, and the latter by the probability measure. It is important to remark that the predictive power of a path space derives from the {\it macroscopic} structure of the trajectories rather than from their microscopic structure. This derives from the fact that our perception of reality is macroscopic, and our knowledge of the microscopic reality is always mediated by macroscopic instruments. For example, a path space correctly predicts the results of quantum experiments if it correctly predicts the evolution of macroscopic apparatus and of their pointers, while the detailed microscopic structure of trajectories is irrelevant from the predictive point of view. This fact has two important consequences: First of all, it implies that the trajectories of a path space need to be only vaguely defined. This allows the path space derived from PSD to make effective predictions, even though it is only vaguely defined, provided the vagueness disappears on a macroscopic scale. The second consequence is that theories defining trajectories with different microscopic structures may be empirically equivalent. As we will see, this is the case for PSD theory and Bohmian mechanics.

PSD theory allows us to derive a path space from the wave function of the universe and therefore has a high predictive power. Note for example that it provides a way to derive the laws of quasi-classical macroscopic evolution from the basic formalism of quantum mechanics, while such a derivation is a well known problem of orthodox quantum mechanics. On the contrary, the explanatory power of PSD appears to be very low. In fact, also due to their vague definition, the trajectories derived by PSD cannot be reasonably considered as the ontological trajectories of the particles, and they are better interpreted as mathematical tools for expressing the predictive power of the wave function. Thus, PSD fails to solve various conceptual problems, such as: why does (our perception of) the evolution of a macroscopic system correspond to only one of the elements of a PSD? Or, what is the primitive ontology of the world, i.e., the stuff that things are made of \cite{Allori:2006}?

Let us compare the predictive powers of PSD and of Bohmian mechanics. More specifically, let us compare the path spaces $\mathscr{D}_\Psi$ and $\mathscr{B}_\Psi$. As mentioned in the introduction, it is a common assumption that Bohmian trajectories remains inside the non-overlapping branches of the wave function. Therefore, the difference between $\mathscr{D}_\Psi$ and $\mathscr{B}_\Psi$ is that in $\mathscr{D}_\Psi$ the Bohmian trajectories contained in a branch are replaced by the trajectory of the centroid of the branch and the same probability of the set of Bohmian trajectories is assigned to it. It is reasonable to argue that this change is not detectable on a macroscopic scale. We are therefore led to the important conclusion that PSD theory has the same predictive power as Bohmian mechanics, even without the guidance equation. On the contrary, Bohmian mechanics is certainly superior to PSD theory with respect to explanatory power. This depends on the fact that the trajectories of $\mathscr{B}_\Psi$ are interpreted as ontological trajectories, i.e., it is assumed that the particles of the universe actually follow a trajectory of $\mathscr{B}_\Psi$. This allows Bohmian mechanics to provide a satisfying answer to the two conceptual questions previously mentioned.

It is well known that the ontological interpretation of Bohmian trajectories excludes a covariant relativistic formulation of this theory. It is useful to recall in general terms why this problem arises, and to show why it is absent in PSD theory. Suppose we have a relativistic quantum theory and a relativistic guidance equation. Under a Poincar\'e transformation the space $\mathscr{B}_\Psi$ can be transformed in the following two ways: (i) the wave function $\Psi$ is transformed to $\Psi'$ and then $\mathscr{B}_{\Psi'}$ is derived from $\Psi'$ by means of the guidance equation or (ii) the trajectories of $\mathscr{B}_\Psi$ are directly transformed as ontological trajectories in space-time. The problem is that, due to Hardy's paradox, there are situations in which these two transformations cannot coincide \cite{Berndl:1995}. This problem does not exist in PSD theory because the trajectories of $\mathscr{D}_\Psi$ are not considered as ontological trajectories and are not required to transform according to an autonomous transformation law.

The formulation (b) in the introduction, based on the Quantum Cournot Principle also contains ontological trajectories. For this reason, the same conclusions relative to Bohmian mechanics arguably hold true for this formulation also, namely: it has the same predictive power as PSD theory, it has an explanatory power superior to that of PSD theory, and it does not admit a covariant relativistic formulation.

\section{Summary} \label{summary}

The focus of this paper is centered on what has been referred to as the {\it permanent spatial decomposition} (PSD) of the wave function, i.e., the (hypothesized) property of the wave function of a macroscopic system of decomposing into permanently non-overlapping parts. This phenomenon is often mentioned in the literature, but until now no formal definition or systematic study of this phenomenon has been undertaken. The aim of this paper is to partially fill this gap by giving a formal definition of PSD and by arguing a connection between this phenomenon and scattering theory.

A generic decomposition of a state vector $\Psi$ has been defined as an unordered set of linearly independent vectors whose sum is $\Psi$. An {\it exact spatial decomposition} is a decomposition of the type $\{E(\Delta_1)\Psi, \ldots, E(\Delta_n)\Psi\}$, where $\{\Delta_1, \ldots, \Delta_n\}$ is a partition of configuration space and $E$ is the spatial spectral measure. An {\it (approximate) spatial decomposition} is a decomposition which is ``close'' to an exact spatial decomposition, where the notion of ``closeness'' is mathematically defined by a specific proximity function $w_E$. Eventually, a {\it permanent spatial decomposition} is a decomposition which remains close to an exact spatial decomposition over the evolution of time. The subsequent occurrence of PSD is represented by a {\it spatial tree}, i.e., a sequence of PDSs, every one of which is ``finer'' than the previous one.

The notions of {\it centroid} and {\it spread} of a state vector have been defined in a general way in terms of a spectral measure on a metric configuration space. These notions allow us to define localized state vectors and to associate a trajectory with every branch of a spatial tree.

The hypothesis of PSD has been precisely reformulated by stating that the wave function of the universe admits a spatial tree whose branches are localized in a time interval which includes the present age of the universe and possibly excludes the initial age of the big-bang and the final asymptotic age.

It has been mathematically proved that, at least in the asymptotic regime, the wave function has a structural tendency to decompose into permanently non-overlaying parts, corresponding to the different scattering channels. This suggests the following representation of the phenomenon of PSD: the elements of a PSD belong to different channels of the wave function of the universe, and the spatial separation of these channels begins long before the wave function enters the asymptotic regime.

Under the assumption that the PSD hypothesis is correct, it has been argued that the {\it predictive power} of PSD, i.e., its capacity for making empirically verifiable predictions, is equivalent to that of Bohmian mechanics, even though the former theory does not include ontological trajectories and the guidance equation. In particular, PSD allow us to predict the structure of macroscopic quasi-classical evolution from the quantum formalism, which is a well known difficulty of standard quantum mechanics. On the contrary, the {\it explicative power} of PSD, i.e., its capacity to explain the nature of physical reality and the origin of observed phenomena, is poor compared to that of Bohmian mechanics. This is due to the absence of ontological trajectories, which prevents PSD from explaining why the (observed) evolution corresponds to only one of the elements of decomposition. On the other hand, the absence of ontological trajectories arguably allows PSD theory to admit a relativistic covariant formulation, which is forbidden in Bohmian mechanics.

\section{Appendix} \label{appendix}
In this section some definitions and theorems relative to spectral measures are presented which are necessary for the proof of theorem \ref{tbase}. Analogous definitions and theorems for scalar measures can be found in textbooks, but this is not the case for those relative to spectral measures.

A spectral measure on a measurable space $(\G, {\cal A})$ is a $\sigma$-additive set function $G$ from ${\cal A}$ to the set of projections of $\HH$ such that $G(\G)=I$. One can prove that a spectral measure is multiplicative, i.e., $G(\Delta_1 \cap \Delta_2)=G(\Delta_1)G(\Delta_2)$ for $\Delta_1, \Delta_2 \in {\cal A}$, and therefore all the projections of a spectral measure commute. All the spectral measures considered here will be defined on the same {\it metric} space $\X=\R^N$, endowed with the Borel $\sigma$-algebra $\BB_\X$. If $f$ is a complex measurable function on $\X$, the integral $\int f dG$ is an operator defined as follows: $\langle \Psi|\int f dG|\Phi \rangle:=\int f dG_{\Psi, \Phi}$, where $G_{\Psi, \Phi}$ is the complex measure $\langle \Psi|G(\cdot)|\Phi \rangle$. If $f$ is real $\int f dG$ is self-adjoint. The positive measure $\langle \Psi|G(\cdot)|\Psi \rangle$ is denoted by $G_\Psi$. $C(\X)$ denotes the class of continuous bounded functions defined on $\X$, and $C_\infty(\X)$ the subset of $C(\X)$ composed of the functions which satisfy $\lim_{x \to \infty} |f(x)|=0$. If a sequence of operators $A_1, A_2, \ldots $ converges strongly to an operator $A$, we write $A_n \to A$.
\begin{lem} \label{l1}
Let $G$ and $G'$ be two spectral measures; then $G=G'$ iff $G_\Psi=G'_\Psi$ for every $\Psi \in \HH$.
\end{lem}
\begin{proof}
The implication of $G=G' \Rightarrow G_\Psi=G'_\Psi$ is obvious. If $G_\Psi=G'_\Psi$ for all $\Psi \in \HH$, by means of the polarization identity we obtain:
$$
\hbox{Re} \langle \Psi |G(\Delta)|\Phi \rangle = \frac{1}{4}[G_{\Psi + \Phi} (\Delta) - G_{\Psi - \Phi} (\Delta)]=
\frac{1}{4}[G'_{\Psi + \Phi} (\Delta) - G'_{\Psi - \Phi} (\Delta)]=\hbox{Re} \langle \Psi |G'(\Delta)|\Phi \rangle,
$$
and analogously $\hbox{Im} \langle \Psi |G(\Delta)|\Phi \rangle=\hbox{Im} \langle \Psi |G'(\Delta)|\Phi \rangle$.
\end{proof}
\begin{thm} \label{t1}
Let $G$ and $G'$ be two spectral measures; if $\int f dG=\int f dG'$ for any $f \in C(\X)$ then $G=G'$.
\end{thm}

\begin{proof}
This theorem is valid for scalar measures \cite{billingconv}. Then $\int f dG_\Psi =\langle \Psi |\int f dG |\Psi \rangle = \langle \Psi |\int f dG' |\Psi \rangle =\int f dG'_\Psi$ for all $\Psi \in \HH$. From the theorem for scalar measures, it follows that $G_\Psi=G'_\Psi$ for all $\Psi \in \HH$, and from lemma \ref{l1} that $G=G'$.

\end{proof}
\begin{defn}
A sequence $G_1, G_2, \ldots$ of spectral measures is said to converge weakly to $G$ if $\int f dG_n \to \int f dG$ for every $f \in C(\X)$.
\end{defn}
If $G_1, G_2, \ldots$ converges weakly to $G$ we write $G_n \to G$. Theorem \ref{t1} guarantees that the limit of a convergent sequence is unique.
\begin{defn}
A {\it $G$-continuity set} is a set $\Delta \in \BB_\X$ such that $G(\partial \Delta)=0$.
\end{defn}
\begin{lem}
The class of $G$-continuity sets is an algebra.
\end{lem}
\begin{thm} \label{t2}
$G_n \to G$ iff $G_n(\Delta) \to G(\Delta)$ for all $G$-continuity sets $\Delta$.
\end{thm}

\begin{proof} See \cite{convvectmes}. \end{proof}
Let $B:=(B_1, \ldots, B_N)$ be a vector of pairwise commuting self-adjoint operators. According to the spectral theorem, this vector univocally defines a spectral measure $G$ on $\X$. If $f$ is a measurable function on $\X$, we define $f(B):=\int f dG$. In \cite{derez} the convergence of vector operators is defined as follows:
\begin{defn} \label{cinf}
A sequence $B_1, B_2, \ldots $ of self-adjoint vector operators is said to converge strongly-$C_\infty$ to $B$ if $f(B_n) \to f(B)$ for all $f \in C_\infty(\X)$.
\end{defn}
If $B_1, B_2, \ldots $ converges strongly-$C_\infty$ to $B$, we write $B_n \Rightarrow B$. This definition of convergence is different from the strong convergence of the single elements of the vectors (the two definitions coincide if the $B_n$ are uniformly bounded in $n$). Due to the way in which the function of a vector operator is defined, the definition of strong-$C_\infty$ convergence for vector operators is almost equivalent to the definition of weak convergence for spectral measures, the only difference being that the function $f$ must belong to $C_\infty(\X)$ in the former and to $C(\X)$ in the latter. Actually, these are equivalent:
\begin{lem} \label{lem1}
$B_n \Rightarrow B$ iff $G_n \to G$, where $G_n$ and $G$ are the spectral measure defined by $B_n$ and $B$, respectively.
\end{lem}
\begin{proof}
It is sufficient to prove that if $f(B_n) \to f(B)$ for all $f \in C_\infty(\X)$ then $f(B_n) \to f(B)$ for all $f \in C(\X)$. See the second part of the proof of proposition 10.1.9 in \cite{intermediate}.
\end{proof}
\begin{thm}
$\BB_\X$ cannot contain an uncountable, disjoint collection of sets of non-null $G$-measure.
\end{thm}
\begin{proof} This theorem holds true for probability measures (see \cite{billingprob}, theorem 10.2), and for any spectral measure $G$ there exists a vector $\Psi$ such $G(\Delta)=0$ iff $G_\Psi(\Delta)=0$ (see \cite{birman}, theorem 7.3.4 pag. 171).
\end{proof}
\begin{thm}
Let $G$ be a spectral measure and $\Delta$ a given measurable subset of $\X$; for any $\Psi \in \HH$ and $\epsilon >0$ there exists a $G$-continuity set $\Sigma$ such that $||G(\Delta)\Psi - G(\Sigma)\Psi|| \leq \epsilon$.
\end{thm}
\begin{proof}
(a) Assume $\Delta$ to be closed. Let $d$ denote the distance on $\X$. Since $\partial \{x \in \X: d(x, \Delta) \leq \delta\}$ is contained in $ \{x \in \X: d(x, \Delta) = \delta\}$, these boundaries are disjoint for distinct $\delta$, hence at most a countable set of them can have non-null $G$ measure. Therefore, for some sequence of positive $\delta_k$ going to $0$, the sets $\Delta_k := \{x \in \X: d(x, \Delta) \leq \delta_k\}$ are $G$-continuity sets. Since $\Delta_k \downarrow \Delta$, it follows that $||G(\Delta_k)\Psi - G(\Delta)\Psi||^2 = G_\Psi(\Delta_k) - G_\Psi(\Delta) \to 0$. (b) Assume that $\Delta$ is not closed. Since $G_\Psi$ is regular, for all $\delta > 0$ there exists a closed set $\Delta' \subseteq \Delta$ such that $||G(\Delta)\Psi - G(\Delta')\Psi||^2 = G_\Psi(\Delta \setminus \Delta') \leq \delta^2$. Let $\Delta'_k$ be a sequence of $G$-continuity sets such that $\Delta'_k \downarrow \Delta'$. Then $||G(\Delta)\Psi - G(\Delta'_k)\Psi|| \leq ||G(\Delta)\Psi - G(\Delta')\Psi|| + ||G(\Delta')\Psi - G(\Delta'_k)\Psi|| \to \delta$. The theorem follows from the fact that $\delta$ is arbitrary.
\end{proof}
\begin{cor} \label{cor1}
Let $G$ be a spectral measure and $\{\Delta_1, \ldots, \Delta_n\}$ a given partition of $\X$; for any $\Psi \in \HH$ and $\epsilon >0$ there exists a partition $\{\Sigma_1, \ldots, \Sigma_n\}$ composed of $G$-continuity sets such that $||G(\Delta_I)\Psi - G(\Sigma_I)\Psi|| \leq \epsilon$ for any $I \subseteq \{1, \ldots, n\}$.
\end{cor}
\begin{proof}
For $i=1, \ldots, n-1$ let $\Delta'_i$ be a $G$-continuity set such that $||G(\Delta_i)\Psi - G(\Delta'_i)\Psi|| \leq \delta$, where $\delta=\epsilon/[2 (n-1)^2]$. For $i=1, \ldots, n-1$ define $\Sigma_i :=\Delta'_i \setminus \bigcup_{j < i} \Delta'_j$, and $\Sigma_n :=\X \setminus \bigcup_{j < n} \Delta'_j $. Then $\{\Sigma_1, \ldots, \Sigma_n\}$ is the required partition. The sets $\Sigma_i$ are $G$-continuity sets. If $i >j$ then $\Sigma_i \cap \Sigma_j = \Delta'_i \bigcap_{k < i} {\Delta'_k}^c \cap \Delta'_j \bigcap_{l < j} {\Delta'_l}^c= {\Delta'_j}^c \cap \Delta'_j \bigcap \ldots = \emptyset$; moreover $\Sigma_n \cup \Sigma_{n-1}= \X \bigcap_{i < n - 1} {\Delta'_i}^c$, $\Sigma_n \cup \Sigma_{n-1} \cup \Sigma_{n-2}= \X \bigcap_{i < n-2} {\Delta'_i}^c$, and so on, thus $\bigcup_n \Sigma_n = \X$. Let us now evaluate $||G(\Delta_i)\Psi - G(\Sigma_i)\Psi||$. For $i < n$ we have:
$$
\Delta'_i - \Sigma_i = \Delta'_i \setminus \Sigma_i= \Delta'_i \cap (\Delta'_i \cap_{j < i} {\Delta'_j}^c)^c = \Delta'_i \cap ({\Delta'_i}^c \cup_{j < i} \Delta'_j)=\cup_{j < i} (\Delta'_i \cap \Delta'_j);
$$
As a consequence
\begin{eqnarray*}
& & ||G(\Delta_i)\Psi - G(\Sigma_i)\Psi|| \leq ||G(\Delta_i)\Psi - G(\Delta'_i)\Psi||+||G(\Delta'_i)\Psi - G(\Sigma_i)\Psi|| \\
& & \leq \delta + ||G[\cup_{j < i} (\Delta'_i \cap \Delta'_j)]\Psi|| \leq \delta + \sum_{j < i} ||G(\Delta'_i)G(\Delta'_j)\Psi||;
\end{eqnarray*}
But
\begin{eqnarray*}
& & ||G(\Delta'_i)G(\Delta'_j)\Psi|| \leq ||G(\Delta'_i)[G(\Delta'_j)-G(\Delta_j)]\Psi|| + ||G(\Delta'_i)G(\Delta_j)\Psi|| \\
& & \leq \delta + ||G(\Delta_j)[G(\Delta'_i)-G(\Delta_i)]\Psi|| \leq 2 \delta.
\end{eqnarray*}
Thus, for $i < n$, 
$$
||G(\Delta_i)\Psi - G(\Sigma_i)\Psi|| \leq \delta + 2(i-1) \delta = (2 i -1) \delta.
$$
Moreover
\begin{eqnarray*}
& & ||G(\Delta_n)\Psi - G(\Sigma_n)\Psi|| = \Big |  \Big | \sum_{i < n}G(\Delta_i)\Psi - \sum_{i < n}G(\Sigma_i)\Psi \Big | \Big | \\
& & \leq \sum_{i < n} ||G(\Delta_i)\Psi - G(\Sigma_i)\Psi|| \leq \sum_{i < n} (2i -1) \delta= (n-1)^2 \delta.
\end{eqnarray*}
Eventually
$$
||G(\Delta_I)\Psi - G(\Sigma_I)\Psi|| \leq \sum_{i \in I} ||G(\Delta_i)\Psi - G(\Sigma_i)\Psi|| \leq 2 (n-1)^2 \delta = \epsilon.
$$
\end{proof}
\bibliographystyle{hplain}
\bibliography{general}
\end{document}